\documentclass[useAMS,usenatbib]{mn2e}
\usepackage{graphicx}

\usepackage{subfigure}
\usepackage{amsmath}

\title[Disc Galaxy Decomposition]{Thin disc, Thick Disc and Halo in a Simulated Galaxy.}
\author[C.B. Brook et~al.]{C.\,B. Brook$^{1,2}$, G.\,S. Stinson$^{3}$, 
B.\,K. Gibson$^{2,4,5}$, D. Kawata$^{6}$, E.\,L. House$^{2}$,  
\newauthor  M.\,S. Miranda$^{1}$, A.\,V. Macci\`o$^{2}$, K. Pilkington$^{2,4,5}$, R. Ro\v{s}kar$^{7}$,
J. Wadsley$^{8}$, T.\,R. Quinn$^{9}$\\
$^1$Departamento de F\'{i}sica Te\'{o}rica, Universidad Aut\'{o}noma de 
Madrid, E-28049 Cantoblanco, Madrid, Spain\\
$^2$Jeremiah Horrocks Institute, University of Central Lancashire, 
Preston, PR1~2HE, UK \\
$^3$Max-Planck-Institut f\"ur Astronomie, K\"onigstuhl 17, 69117 
Heidelberg, Germany\\
$^4$Department of Astronomy \& Physics, Saint Mary's University, 
Halifax, Nova Scotia, B3H~3C3, Canada\\
$^5$Monash Centre for Astrophysics, Monash University, VIC 3800, Australia\\
$^6$Mullard Space Science Laboratory, University College London, 
Holmbury St. Mary, Dorking, Surrey RH5 6NT\\
$^7$Institute for Theoretical Physics, University of Z\"{u}rich, Winterthurerstrasse 190, CH-8057 Z\"{u}rich, Switzerland\\
$^8$Department of Physics and Astronomy, McMaster University, Hamilton, 
Ontario, L8S~4M1, Canada\\
$^9$Astronomy Department, University of Washington, Box 351580, Seattle, 
WA 98195-1580, USA \\
}

\begin{document}

\date{}

\pagerange{\pageref{firstpage}--\pageref{lastpage}} \pubyear{2011}

\maketitle

\label{firstpage}

\begin{abstract}
Within a cosmological hydrodynamical simulation, we form a disc galaxy 
with sub-components which can be assigned to a thin stellar disc, 
thick disk, and a low mass stellar halo via  a 
chemical decomposition. The thin and thick disc populations so selected are distinct 
in their ages, kinematics, and metallicities. Thin disc stars are young 
($<$6.6~Gyr), possess low velocity dispersion ($\sigma_{\rm U,V,W}=41,31,25$~km\,s$^{-1}$),  high [Fe/H], and low 
[O/Fe]. Conversely, the thick disc stars are old (6.6$<$age$<$9.8~Gyrs), 
lag the thin disc by $\sim$21~km/s, possess higher velocity dispersion 
($\sigma_{\rm U,V,W}=49,44,35$~km\,s$^{-1}$), and have
relatively low [Fe/H] and high [O/Fe]. The halo component comprises 
less than 4\% of stars in the ``solar annulus'' of the simulation, has 
low metallicity, a velocity ellipsoid defined by ($\sigma_{\rm U,V,W}=62,46,45$~km\,s$^{-1}$) 
and is formed primarily \it in-situ \rm during an early merger epoch. 
Gas-rich mergers during this epoch play a major role in fuelling the 
formation of the old disc stars (the thick disc). We demonstrate 
that this is consistent with studies which show 
that cold accretion is the main source of a disc galaxy's baryons. Our 
simulation initially forms a relatively short (scalelength $\sim$1.7\,kpc
at $z$=1) and kinematically hot disc, primarily from gas accreted during 
the galaxy's merger epoch. 
Far from being a competing formation 
scenario, we show that migration is  crucial for reconciling the
short, hot, discs which form at high redshift in 
$\Lambda$CDM, with the properties of the thick disc at $z$=0.
The thick disc, as defined by its abundances maintains its relatively short scale-length at 
$z=0$ (2.31\,kpc) compared with the total disc scale-length of 2.73\,kpc.
The inside-out nature of disc growth is imprinted the evolution of abundances such that  the metal poor $\alpha$-young
population has  a larger scale-length (4.07\,kpc) than the more chemically evolved metal rich $\alpha$-young population (2.74\,kpc).

\end{abstract}


\section{Introduction}
\label{introduction}

 The fitting of the vertical distribution by two exponentials has motivated separation of the Milky Way's disc population into   
 `thick' and `thin' stellar disc components. Such classification has dominated the 
literature of Galactic structure for the past three decades 
\cite[e.g.][although see \citealt{ivezic08,bovy12a}]{gilmore83,juric08}. 
The oldest disc stars in the Milky Way are kinematically hot, show an 
asymmetric drift (i.e. rotational lag), and have lower metallicities 
([Fe/H]) and higher $\alpha$-to-iron abundance ratios, 
with respect to the younger disc stars.
Understanding the origin of the stars distributed at the highest 
vertical scaleheights within the disc - the classical `thick disc' - is 
an important aspect in our attempts to understand galaxy formation and 
evolution throughout the Universe, as such thick discs appear ubiquitous 
\citep[e.g.][]{dalcanton02,yoachim05}, and contain the oldest disc stars, born during the earliest formation of discs.

Several models for the formation of such thick stellar discs have been 
proposed, each of which are able to reproduce one or more of their 
empirical properties. Despite these partial successes, a singular, 
universally accepted model has not yet emerged.  Such a model needs to 
answer two fundamental points: (i) the origin of the oldest disc stars, 
and (ii) how these stars attained their present day properties.

The heating of a pre-existing thin disc due to the accretion of 
satellite galaxies \citep{quinn93} appears consistent with the 
hierarchical structure formation of $\Lambda$CDM.  The signatures of 
such events are well studied and indeed are able to match vertical 
profiles of the thick disc 
\citep{hayashichiba06,stelios08,villalobos08,qu11,bekki11}, including 
several fine-detailed features \citep{qu11}. Such models assume a 
pre-existing and relatively thin disc.  Another model proposes 
that stars accreted by satellites can be dragged into the plane of a 
pre-existing disc by dynamical friction \citep{abadi03b}. Recently, it 
has been shown that the radial migration of old, $\alpha$-enhanced stars 
from the inner region of the disc to the solar neighbourhood can explain 
many properties of the thick disc \citep{sb09b,loebman11}.  The ``popping" of star clusters 
\citep{kroupa02a,assmann11} or the effects of massive clump formation in 
unstable gas-rich discs in the early universe 
\citep{noguchi99,bournaud07,agertz09,ceverino10} may also result in thickening of 
discs; such scenarios link thickening more to formation processes, 
rather than to disc heating, per se. Stars may also have been born 
relatively thick as gas settles into a disc configuration during an 
early period of gas-rich mergers \citep{brook04a}. 

It is important to stress that these models are not necessarily mutually 
exclusive. Indeed, some do not try to explain the formation of the 
oldest stars, but rather how they were heated, or how they migrated to 
the solar region. One way to comprehensively probe thick 
disc formation is using high-resolution, hydrodynamical, cosmological 
galaxy formation simulations of Milky Way mass galaxies 
\citep{abadi03b,stinson10,scannapieco10,kobayashi11,tissera12,martig12,domenech12}. 
These attempts, though, have been hampered by their inability to match 
several important properties of the Milky Way. In this context, the main 
problems are (i) the continued formation of overly massive and metal 
rich stellar halos, which makes separation of thick disc and halo stars 
more difficult than in the Milky Way where metallicity can be used as an 
added discriminant, and (ii) the lack of resolution. The first problem 
has been alleviated by improved feedback (\citealt{brook04b}; \citealt{okamoto05,g07,guedes11}) 
but continues to plague even  recent simulations 
\citep{sales12,scannapieco12,martig12}. The latter problem is more 
subtle than just the size of gravitation softening, but relates to the 
relatively low density and high temperature at which star formation 
occurs in low-resolution simulations, which prevents thin discs from 
forming \citep{house11}.

In this study, we take a complementary approach by examining a disc 
galaxy simulation of lower mass than the Milky Way, providing us with 
improved spatial resolution and allowing stars to form at higher 
densities and lower temperatures. We analyse the simulation first 
presented in \citet[][B12, hereafter]{brook12a} 
late-type barred galaxy with M$_{vir}$=1.9$\times$10$^{11}$\,M$_\odot$, 
realised as part of the MaGICC (Making Galaxies In a Cosmological 
Context) program. To re-iterate, this simulation is of a lower mass 
than that of the Milky Way, and therefore our analysis must be read 
within this context. That said, we note that the simulated galaxy 
examined here shares other properties with the Milky Way that we feel 
make it relevant to analyse in relation to putative theories of thick 
disc formation. Although the Milky Way is the only galaxy for which we 
have detailed abundances and kinematic information, vertically-extended 
`thick' discs appear ubiquitous in nature. Any comparison we make with 
the Milky Way is not expected to match \it exactly\rm, but is predicated 
upon the fact that the broad trends of the Milky Way chemo-dynamics 
should be generic for relatively massive disc galaxies {\it which have 
quiescent merger histories}, in the sense that the last major merger was 
early in the galaxy's evolution. There is evidence that the Milky Way 
had such a quiescent merging history \citep[e.g.][]{hammer07}. However, 
if these assumptions were to be shown to be false, our model would 
clearly unravel.

The second  feature of our study is our chemical decomposition of 
stars into different galactic sub-components based on abundance 
patterns.  The ability to separate components based purely on their 
chemical ``tags" is at the heart of Galactic Archeaology 
\citep{freeman02}.  A multitude of studies has shown that  thin and thick disc stars 
with similar iron abundances ([Fe/H]) , differ in their abundances of $\alpha$ elements \citep[e.g.][]{majewski93,fuhrmann98,chibabeers00,bensby03,soubiran03,wyse06,reddy06,fuhrmann08,ruchti10}.
Using a compilation of observational data, 
\cite{navarro11} made simple cuts in the [$\alpha/$Fe]-[Fe/H] plane, in 
order to separate the Milky Way's thick and thin discs. 
We  adopt a similar decomposition in the simulated [O/Fe]-[Fe/H] 
plane. \cite{bensbyfeltzing11} argue that the large overlap in 
kinematics between thick and thin disc populations makes classification 
based on age better than classification based on kinematics. The 
relative difficulty of ascertaining stellar ages means that abundances, 
in conjunction with abundance ratios, may be the best ``proxy''. \cite{bovy12b} make detailed cuts in abundance space 
to probe thick disc properties.

\begin{figure}
\includegraphics[width=.32\textheight]{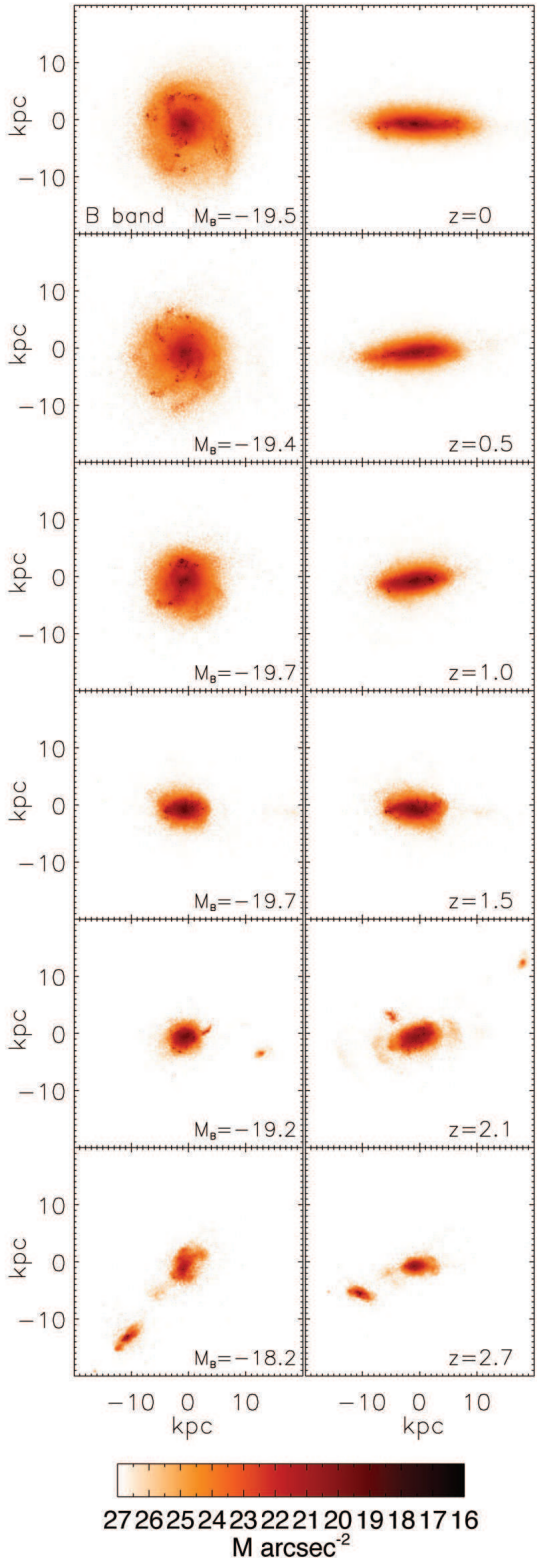}
\caption{The evolution of the simulated galaxy shown in $B$-band surface 
brightness maps, as viewed both face-on (left panels) and edge-on (right 
panels). The redshifts range from $z$=2.7 to $z$=0, and are noted in the 
bottom right corner of the edge-on panels. Each panel is 40$\times$40\,kpc (physical units). The effects of dust reprocessing, as implemented with 
\textsc{Sunrise}, are included.}
\label{evolve}
\end{figure}

Thirdly, our study proposes more detailed categorisations of the origin 
of stars that comprise the different components, going beyond {\it accreted} 
vs \it in situ\rm. In particular, we concern ourselves with whether gas 
which subsequently forms stars has been involved in a ``gas-rich 
merger'', as well as whether it was smoothly accreted or accreted 
directly from proto-galaxies/satellites. This provides a more complete 
perspective of the relative importance of gas-rich mergers compared with 
smooth  accretion.

In Section~\ref{code}, we review our code and initial conditions, and 
outline the basic properties of the resulting simulation. In 
Section~\ref{properties}, we follow \cite{navarro11} and separate our 
(analogous) solar neighbourhood stars on the basis of their abundances, 
and find that such classification results in components that 
could  reasonably be referred to as a thin disc, thick disc, and 
halo/metal weak thick disc, sharing some gross characteristics to those 
in the Milky Way. In Section~\ref{origins} we trace the origin of the 
stars in these populations, both in terms of where they formed (\it in 
situ \rm or {\it accreted}, and where in the disc that they formed), their 
formation dispersion, and the origin of the gas from which they formed. 
We find that thick and thin disc stars form \it in situ \rm and that 
cold accretion and gas-rich mergers supply thick disc gas. We 
find in Section~\ref{migration} that migration also plays a role in forming the 
simulated thick disc.  In Section~\ref{discussion}, we discuss the 
consistency of these results and the fact that most disc stars form from 
smooth accretion and highlight the role of migration in 
shaping our thick disc population.

\section{The Simulation}
\label{code}
\subsection{Code and Initial Conditions}
We analyse the simulation described in B12. 
 To quickly review, this 
is  a cosmological `zoom' simulation, derived from the initial 
conditions associated with galaxy \tt g15784 \rm from the McMaster 
Unbiased Galaxy Simulations (MUGS, \citealt{stinson10}), using a 
25~$h^{-1}$~Mpc parent cube. The virial mass of the simulated galaxy is 
M$_{vir}=1.94\times10^{11}$~M$_{\odot}$. Further detailed chemical 
modelling (see below) has been incorporated into the code, and the 
simulation re-run: the stochastic nature of our star formation recipes 
mean that small differences exist between the current version of the 
simulated galaxy and the one analysed in B12, but essential properties 
such as rotation curve shape, star formation history, magnitude, colour, 
and disc scalelength are not altered significantly. In B12, we showed 
that secular processes, largely driven by a strong bar, resulted in the 
formation of a bulge with B/T=0.21 at $z$=0. In the current simulation, 
a strong bar does not form and hence, neither does such a significant 
bulge. The current simulation has B/T=0.12 at $z$=0.

\begin{figure*}
\includegraphics[width=.47\textheight]{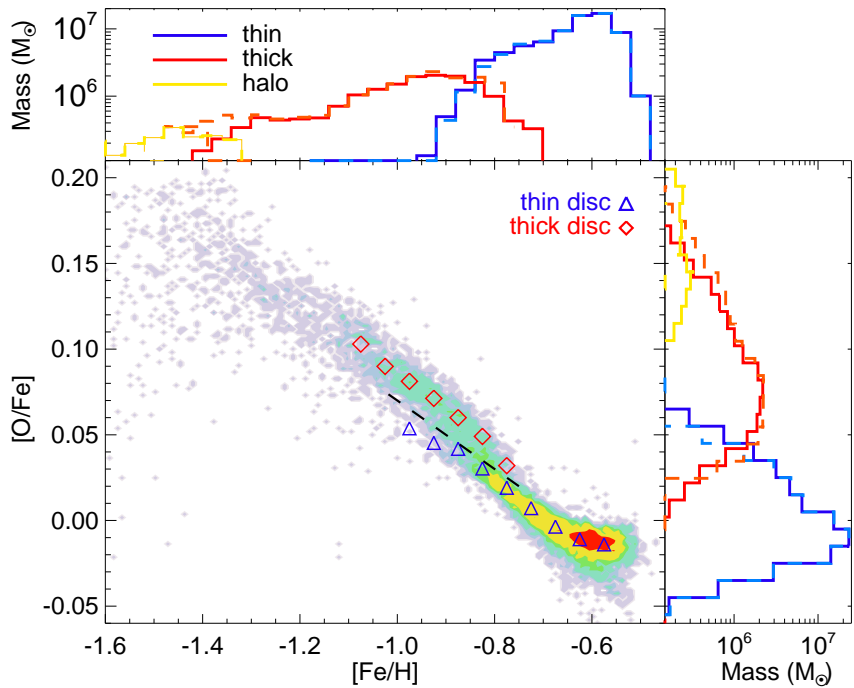}
\caption{[O/Fe]-[Fe-H] of simulation stars in the region $7<Rxy<8$\,kpc, 
$|z|<0.5$kpc. Two ``evolutionary tracks'' are evident, loosely 
delineated by the dashed line: [O/Fe]=-0.24$\times$[Fe/H]$-$0.16. This 
simple delineation defines what we label `thick' and `thin' disc 
populations in this work. We then overplot the means of the thick  (red 
diamonds) and thin (blue triangles) discs [O/Fe] as a function of 
[Fe/H], for the populations as defined by their ages (see text).
Above (on the right)  is plotted (on a log scale) the distribution of [Fe/H]  ([O/Fe]) in the `solar annulus' for thin disc 
(blue line), and thick disc (red line) stars, based on abundances (solid) 
and age (dashed).}
\label{ofefeh}
\end{figure*}

The simulation is evolved with the smoothed particle hydrodynamics (SPH) 
code \textsc{gasoline} \citep{wadsley04}.  \textsc{gasoline} employs 
cooling based on the radiative transfer treatment within \textsc{cloudy} 
\citep{shen10}, including cooling due to H, He, a variety of metal 
lines, in the presence of an external UV radiation field. A pressure 
floor is added \citep{robertson08}, and a maximum density limit is 
imposed by setting a minimum SPH smoothing length, $h_{sm}$ to 
$\frac{1}{4}$ of the gravitational softening to ensure that gas resolves 
the Jeans mass and does not artificially fragment. We have $\sim$ 4 
million resolution elements (gas+dark matter+star) within the virial 
radius at $z=0$, with mean stellar particle mass of 7300~M$_{\odot}$ and 
a gravitational smoothing length of 155~pc.

When gas reaches cool (T$<$10,000~K) temperatures in a dense 
($n_{th}$$>$9.3~cm$^{-3}$) environment, it becomes eligible to form 
stars.  This value for $n_{th}$ is the maximum density gas can reach 
using gravity.  Gas is converted to stars according to the equation
\begin{equation}
\frac{\Delta M_\star}{\Delta t} = c_\star \frac{M_{gas}}{t_{dyn}}
\end{equation}
where $\Delta M_\star$ is the mass of the star particle formed, $\Delta 
t$ the timestep between star formation events, $M_{gas}$ the mass 
of the gas particle and $t_{dyn}$ the gas particle's dynamical time.  
$c_\star$ is the efficiency of star formation; in other words, the 
fraction of gas that will be converted into stars during $t_{dyn}$.

Two types of energetic feedback are considered: supernovae and stellar 
radiation.  Supernova feedback is implemented using the blast-wave 
formalism described in \citet{stinson06} and each supernova is assumed 
to deposit $10^{51}$~erg of energy into the surrounding interstellar 
medium (ISM).  Since this gas is dense, the energy would be quickly 
radiated away due to the efficient cooling in SPH.  For this reason, 
cooling is disabled for particles inside the blast region. We model the 
luminosity of stars using the \citet{torres10} mass-luminosity 
relationship. These photons do not couple efficiently with the 
surrounding ISM \citep{freyer06}.  We thus want to couple only a small 
fraction of this energy to the surrounding gas in the simulation. To 
mimic this highly inefficient energy coupling, we firstly inject 10\% of 
the energy as thermal energy in the surrounding gas, and cooling is 
\emph{not} turned off for this form of energy input. It is well 
established that such thermal energy injection is highly inefficient at 
the spatial and temporal resolution of the type of cosmological 
simulations used here \citep{katz92}. This is primarily due to the 
characteristic cooling timescales in the star forming regions being 
lower than the timestep of the simulations.

Metals are ejected from type II supernovae (SNeII), type Ia supernovae 
(SNeIa), and the stellar winds driven from asymptotic giant branch (AGB) 
stars. The ejected mass and metals are distributed to the nearest 
neighbour gas particles using the smoothing kernel \citep{stinson06}, 
and we employed standard yields from the literature for SNeII 
\citep{woosley95} and SNeIa \citep{nomoto97}. Metal diffusion is also 
included, such that unresolved turbulent mixing is treated as a 
shear-dependent diffusion term \citep{shen10}. This allows proximate gas 
particles to mix their metals. Metal cooling is calculated based on the 
diffused metals.

\subsection{Evolution and Properties}
\label{properties}
We emphasise again that this is not a simulated 
Milky Way galaxy. Our simulated galaxy is a less massive late-type 
galaxy, with rotation velocity of $\sim$140~km/s, and a total stellar 
mass within the virial radius of $\sim$8$\times$10$^9$~M$_{\odot}$. The galaxy has an absolute 
magnitude of M$_{B,V,I}$=$-$19.45, $-$19.94, $-$20.71, and sits in the 
blue sequence of the colour-magnitude diagram of observed galaxies.

We plot the morphological evolution of the central galaxy in the $B$-band in 
Figure~\ref{evolve}, face-on (left panels) and edge-on (right panels) 
from $z$=2.7 to $z$=0 (bottom to top in the figure).  Each panel is 40$\times$40 
kpc (physical units).  The major merging 
activity of the galaxy occurs between $z$=2.7 and $z$=1.7, and we refer 
to this period as the ``merger epoch''. Three significant mergers occur 
during this period - firstly, a 3:1 major merger, followed by mergers 
with mass ratios of 10:1 and 20:1, with the  last vestiges of a 
merging satellite still apparent at $z$=1.5.  In B12, we showed that 
over 90\% of the gas which cools to the bulge region during the merger 
epoch is blown out to the hot halo or beyond, rather than forming bulge 
stars. The galaxy does not form a significant bulge during the merger 
epoch and has an almost pure exponential profile at $z$=1 (Sersic index 
$n$=1.1), at which time it has a scalelength of 1.7\,kpc. The 
scale-length of the disc component at $z$=0 is 2.7\,kpc, as measured in 
the $I$-band.


\section{Results}
\subsection{Components}
\label{cuts}

In Fig~\ref{ofefeh}, we plot [O/Fe] vs [Fe/H] in the `solar annulus', 
taken as the region $7<R_{xy}<8$\,kpc and $|z|<0.5$kpc, where the disc is in the $xy$ plane.\footnote{We did similar 
analyses for various regions between $5<R_{xy}<9$\,kpc and $|z|<2$\,kpc; none of the trends discussed are dependent upon the precise 
region chosen as  representative of the disc.} We analyse here oxygen as a representative $\alpha$ element which is synthesised predominantly in Type II SNe, and which is an element whose synthesis is relatively well studied.
An interesting 
feature of this plot is that there appears to be two trajectories, one 
stretching from ([Fe/H],[O/Fe])$\sim(-1.1,0.11)$ to 
([Fe/H],[O/Fe])$\sim(-0.82,0.05)$, and the other running from 
([Fe/H],[O/Fe])$\sim(-0.85,0.035)$ to ([Fe/H],[O/Fe])$\sim(-0.5,-0.02)$,
offset by $\sim$0.01-0.05~dex in [O/Fe]. 
We explore these trajectories by separating them into two populations 
simply using a straight line with equation 
[O/Fe]=-0.24$\times$[Fe/H]$-$0.16 (the dashed line in 
Fig~\ref{ofefeh}).  

\begin{figure}
\hspace{-.5cm} 
\includegraphics[height=.23\textheight]{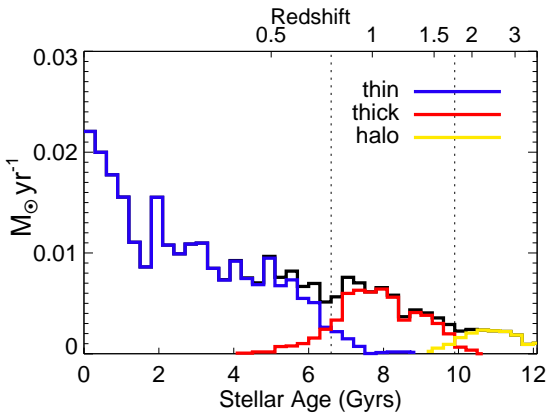}
\caption{The star formation history (SFH) of stars in the solar annulus, defined as the region 
$7<R_{xy}<8$ and $|z|<0.5$\,kpc (black line). The SFH of the thin disc 
(blue line), thick disc (red line), and halo (yellow line), as defined 
by abundances in the [O/Fe]-[Fe-H]  plane, are overplotted.}
\label{tform}
\end{figure}

The $\sim$0.01-0.05~dex vertical offset between the two `tracks' appears to be the result of the reduction 
in the star formation rate (SFR) $\sim$6.6~Gyrs ago, when [Fe/H]$\sim$$-$0.85 (see Figs.~\ref{sfr_rad}\,\&\,\ref{tform}).  
 In the case of the Milky Way, the tracks are offset 
vertically by $\sim$0.2~dex, rather than the $\sim$0.01-0.05~dex seen here.
Yet the existence of these two tracks and the nature of their offset motivates us to  associate the
 populations with the thick (above the line) and thin (below the 
line) discs.

The larger offset in the Milky Way (0.2 dex) has been  interpreted as indicating 
a  hiatus in star formation  in the Milky Way, along with an initiation of  metal-poor gas inflow 
associated with the formation of the thin disc
\citep[e.g.][]{chiappini01,fenner03}. Increased  inflows of metal-poor gas 
are not seen $\sim$6.6~Gyrs ago in this simulation, and we see a star formation reduction rather than hiatus. This could explain the difference with the dual infall models
 in the size of the offset. Further simulations with more massive galaxies will be needed to explore whether a star formation reduction coupled with the gas cycle produced in our simulation can result in larger offsets in the tracks, or whether further processes can be invoked that would lead to a hiatus in star formation.
  We note that  \cite{sb09b} attain  bimodal distributions in [O/Fe] as a  consequence of their  assumptions about star-formation rates and metal enrichment,  without a star formation hiatus. We also repeat the warning provided in \cite{sb09b}: it is important to bear in mind that our simulation data is a kinematically unbiased sample, while most similar observational plots are for samples that are kinematically biased in favour of Òthick-discÓ stars. In fact, \cite{bovy12a} shows that the bi-modality in [$\alpha$/Fe] is greatly diminished once the populations are correctly weighted. 

\begin{figure}
\hspace{-.4cm} 
\includegraphics[height=.28\textheight]{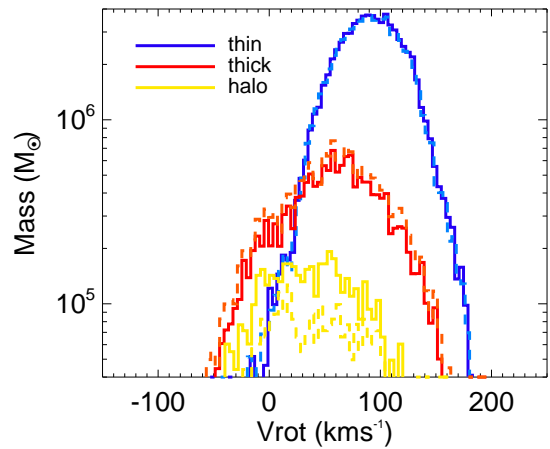}
\caption{The distribution of rotational velocities in the solar annulus 
for the thin disc (blue line) and thick disc (red line) stars, based on 
abundances (solid) and ages (dashed). Note the log scale.}
\label{vrot}
\end{figure}



We next analysed the properties of the two populations of stars as 
defined by the aforementioned linear separation. In Fig~\ref{tform}, we 
plot the star formation history of stars in the solar annulus (black 
line); the formation history of stars above the dashed line in Figure~\ref{ofefeh} (the putative thick disc) are plotted in red 
while the stars below the  dashed line in Figure~\ref{ofefeh} (thin disc) are plotted in blue. 
 A simple cut at the low metallicity end of the thick disc 
trajectory through this abundance space also allows us to classify stars 
with [Fe/H]$<$$-$1.3 as belonging to the stellar halo, although we will 
see that this population may be better described as including a metal 
weak thick disc (MWTD), particularly in the range $-$1.5$<$ [Fe/H]$<$$-$1.3.
The star formation 
history of  halo stars is plotted in yellow in Figure ~\ref{tform}. 
The separation of the ages of stars, as selected by abundances, is 
clear.  

The thin disc stars are 
exclusively young, mostly less than 6.6~Gyr old, but with a small number 
of stars as old as 8.5~Gyr. Thick disc stars are mostly older than 
6.6~Gyr, and range from 5.5 to 10~Gyr. 
The ages of our selected halo 
population ranges from 9.5 to 12~Gyr.  The mass of stars in the thin, 
thick and halo/MWTD components in this region are 22.6, 8.9, and 1.3 
$\times10^7$M$_{\odot}$, respectively. Dotted lines indicate times 
chosen for a pure age classification of the components, which will be 
explored alongside the metallicity cuts in the next sub-section.

\subsection{Component Abundances}

In Figure~\ref{ofefeh} we show the the distributions of 
[Fe/H] and [O/Fe] in the solar annulus stars for the thin disc (blue 
line), thick disc (red line), and halo (yellow) populations, again using 
abundance cuts (solid lines) and age cuts (dashed lines). This merely 
confirms that the selection of stars based on metallicity corresponds 
closely to the ages of the populations.

\subsection{Component Kinematics}

In Fig~\ref{vrot}, we plot the distribution of the rotational velocity 
(V) of the solar annulus stars for the thin disc (blue line) thick disc 
(red line), and halo (yellow) populations, using abundance cuts (solid 
lines) and age cuts (dashed lines) as defined in Section~\ref{cuts}. The 
thick disc lags the thin disc by $\sim$25~km/s. The V distribution of 
halo stars shows there is rotation in this low-metallicity stellar 
population, although there is a hint of more than one population. The 
dispersion in U (radial velocity), V, and W (the velocity perpendicular 
to the disc plane) of the stars in the solar annulus for the thin disc, thick disc and halo, 
 are (in km/s) {\Large $\sigma$}$_{\rm U,V,W}$= 
(41,31,25), (49,44,35) and (62,46,45), respectively.


\begin{figure}
\hspace{-.7cm}  
\includegraphics[height=.22\textheight]{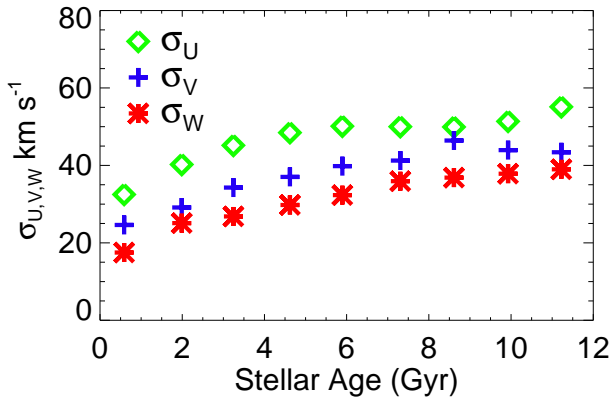}
\caption{The age-velocity dispersion relation for solar annulus stars, 
showing {\Large $\sigma$}$_{\rm U,V,W}$ as green diamonds, blue crosses, 
and red asterisks, respectively.}
\label{sigma_age}
\end{figure}

\subsection{Evolution of Velocity Dispersion}
The age-velocity dispersion relation for solar annulus stars is plotted 
in Fig~\ref{sigma_age}, showing {\Large $\sigma$}$_{\rm U,V,W}$ as green 
diamonds, blue crosses, and red asterisks, respectively. A gradual 
heating is apparent to a lookback time of $\sim$11~Gyr, more in line 
with the conclusions of \cite{nordstrom04} than \cite{quillen01}, 
although this is a complex issue that we have explored earlier in the 
context of more massive disc simulations \citep{house11}.  In this 
latter work, we showed that numerical heating is not causing the 
gradual heating that we are seeing in these simulations.

\begin{figure}
\hspace{-.7cm}  
\includegraphics[height=.22\textheight]{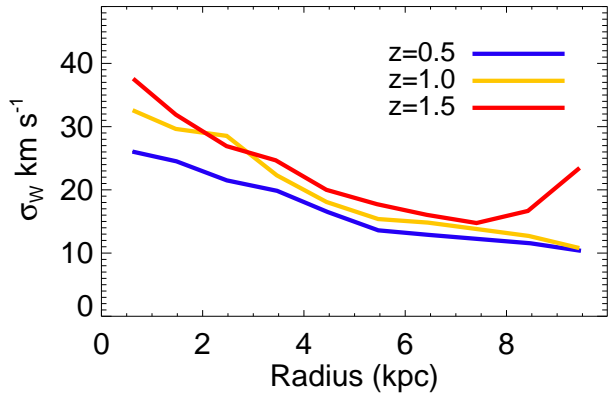}
\caption{ {\Large $\sigma$}$_{W}$ versus radius of young stars 
(age$<$200\,Myrs) in the disc plane ($|z|<$0.5kpc) at three different 
epochs: $z$=1.5 (red line); $z$=1 (yellow line); $z$=0.5 (blue line).}
\label{sigma_radius}
\end{figure}

We examine the evolution of velocity dispersion in more detail in 
Fig~\ref{sigma_radius}, where we plot {\Large $\sigma$}$_{W}$ versus 
radius of young stars (age$<$200\,Myrs) at three different times, 
firstly at $z=1.5$ (red line), a time corresponding to the early period 
of thick disc star formation, then at $z=1$ (yellow line) which is at 
the end of the thick disc star formation, and finally a later time 
(z=0.5 blue line) during the thin disc formation epoch. At each epoch, 
we use young stars in the disc plane ($|z|<$0.5\,kpc) of the galaxy. It is 
evident that the disc stars that were forming at high redshift were born 
kinematically hotter than than those born at later times. Combined with 
Fig~\ref{evolve}, we can say that the stars forming at the end of the 
gas rich merger epoch create a relatively short and kinematically hot 
disc structure. We note that the extent of this old disc is also 
reflected in the $z$=1.5 dispersion-radius plots, with the jump in 
dispersion at R$_{xy}$$\sim$7.5\,kpc., while the disc extends beyond 
9~kpc by $z$=1 (and beyond 12\,kpc at $z$=0). Yet the difference in dispersion at time of birth is not enough to 
explain the  dispersion-age relation in Figure~\ref{sigma_age}. The dispersion radius relation is 
important in this respect. We will see that old stars  in the solar annulus that have large dispersions 
in Figure~\ref{sigma_age}  were preferentially born at lower radii, and hence with relatively high dispersion.

\section{Origin of the Components}\label{origins} 
 33.1$\times$10$^7$M$_\odot$ of stars are in the solar annulus as defined as the region $7<R_{xy}<8$\,kpc and $|z|<0.5$\,kpc. Based on our abundance cuts, 
22.6$\times$10$^7$M$_\odot$ (68\%) are classified as thin disc, 8.9$\times$10$^7$M$_\odot$ (27\%) as thick disc and 
 1.3$\times$10$^7$M$_\odot$ (4\%) as halo/MWTD.
We now trace the gas from which stars in the solar annulus were formed. 
We identify five categories (Table~\ref{origin}) for the origin of the 
stars, based on where they were formed and the source of the gas from 
which they were formed. Firstly, stars formed in proto-galaxies {\it 
other} than the central galaxy and then accreted are classified as {\it 
stellar accretion} (ST ACC). These accreted stars comprise just 0.5\% of stars in 
the solar annulus, and all were classified as halo stars due to their 
abundances being [Fe/H]$<$$-$1.3. So 17\% of the stellar halo in the 
solar annulus was accreted in this manner. This would appear to be a 
fairly low fraction, and we have already stated that the halo as defined 
may include a metal weak thick disc. We make a stricter criteria and 
check {\it retrograde} halo stars, and find that still only 21\% are 
direct {\it stellar accretion}.  We note that accreted stars have metallicities [Fe/H]$<1.8$, so
we finally make an even more strict criteria by limiting our sample to those with [Fe/H]$<1.8$ 
and find that 31\% of stars with these low metallicities are directly accreted.
No disc stars (thick or thin) in our 
simulation have been accreted directly.

Stars forming from gas which is in proto-galaxies {\it other than} 
the central galaxy when it is accreted, but that does not form stars until {\it after} the 
accretion, are  categorised as {\it clumpy gas accretion} (CLPY GAS ACC). 28\% of halo 
stars, 17\% of thick disc stars, 10\% of thin disc stars, and 13\% of 
all stars in the solar annulus were formed via this mechanism.

Stars which formed from gas that was originally accreted smoothly 
but form stars in the central galaxy during the ``merger epoch'', 
were classified as {\it in-situ stellar merger} (I-SU ST MGE) stars: 47\% of halo 
stars and 9\% of thick disc stars are classified as due to an {\it in 
situ star merger}. Stars which form from gas that is originally accreted 
as smoothly to the central galaxy, but do not form stars until 
subsequent to the ``merger epoch'', are classified as {\it in situ gas 
merger} (I-SU GAS MGE) stars: 7\% of halo stars, 49\% of thick disc stars, and 28\% of 
thin disc stars are {\it in situ gas merger} stars, with such stars 
comprising a total of 33\% of all solar annulus stars.  The remaining 
stars form from gas that is smoothly accreted to the central galaxy 
 and are classified as being due to {\it smooth accretion} (SMTH ACC).

The halo component is dominated by  a combination of direct stellar accretion, and 
stars which form in the central galaxy (\it in situ\rm) prior to the end 
of the ``merger epoch'' and are ``knocked'' into the halo by the merger 
events. A significant number of the latter halo stars formed from gas 
that was originally accreted as gas from non-central progenitors. The 
remaining 7\% of the halo stars, as selected by their low metallicity, 
formed from gas which was in the central galaxy during the gas-rich 
mergers but formed stars during the turbulent period which followed the 
merger epoch. These stars actually have a net average rotation of 
30~km/s, and may be associated better with a metal weak thick disc.

\begin{table}

\begin{tabular}{ccccccc}
\hline
&MASS& ST & CLPY & I-SU & I-SU & SMTH\\  
&& ACC &GAS  & ST & GAS & ACC \\  
&(10$^7$M$_\odot$)&   &  ACC& MGE & MGE & \\  
\hline
Total &33.1&0.7\%&18\% & 4.\% &33\% &44\%\\ 
\hline
Thin &22.6&0\% &16\% &0\%& 28\% &56\%\\ 
\hline
Thick &8.9&0\%&23\% & 9\%& 49\% &19\%\\
\hline
Halo &1.3 &17\%&28\%& 47\%& 7\%&1\%\\
\hline
\end{tabular}
\caption{The origin of stars in the solar neighbourhood. {\it Stellar 
accretion} (ST ACC): stars accreted from merged non-central 
progenitors/satellites.  {\it Clumpy Gas Accretion} (CLPY GAS ACC):  gas accreted from 
merged non-central progenitors/satellites.  {\it In-situ Star Merger} (I-SU ST MGE): 
stars born in central galaxy during the merger epoch.  {\it In-situ Gas 
Merger} (I-SU GAS MGE): gas in central galaxy during the merger epoch. {\it Smooth 
Accretion} (SMTH ACC): gas accreted subsequent to the merger epoch.}
\label{origin}
\end{table}

\begin{figure}
\hspace{-.1cm} \includegraphics[height=.25\textheight]{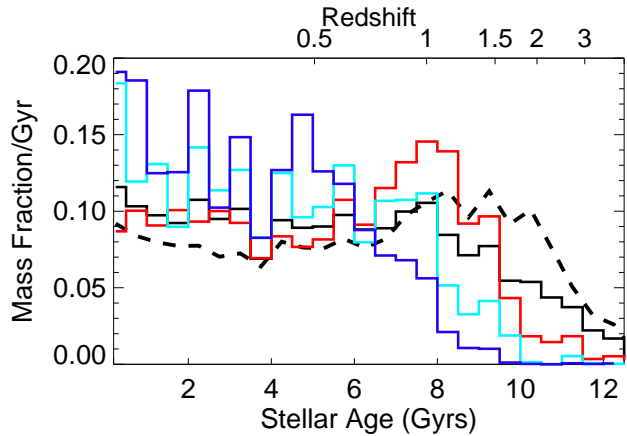}
\caption{The star formation history of all stars of the simulated galaxy 
is shown as the dashed black line, plotting the fraction of stars born per Gyr 
in bins of 0.5~Gyr. Overplotted is the star formation history of stars 
in the disc (black  line). We then show the star formation history 
of disc stars according to different formation radii: 3$<$R$_{\rm Form}$$<$5 
(red), 5$<$R$_{\rm Form}$$<$7 (light blue), and 
7$<$R$_{\rm Form}$$<$9 (blue).}
\centering
\label{sfr_rad}
\end{figure}

What we call `thick disc stars' are those which formed predominantly 
from gas involved in gas-rich mergers, with 49\% being in the central galaxy 
prior to the end of the merger epoch and 17\% in the smaller 
proto-galaxies accreted during this merger epoch. However, the thick 
disc stars which form from this gas do so {\it subsequent} to this 
merger epoch, while the galaxy is `settling'.  Only $\sim$20\% of the 
thick disc stars in the simulation are formed from smoothly accreted, 
non-merger, gas - i.e., gas accreted to the galaxy that is not involved 
in merger events. No thick disc stars are directly accreted during 
mergers. Our findings highlight that gas-rich mergers are a dominant 
process in the formation of thick disc populations. Note that this is 
not at odds with studies which claim that cold flow accretion is the 
dominant source of a galaxies baryons. Rather, these studies, as well as 
ours, need to be interpreted carefully: baryons accreted prior to a 
major merger in studies of cold accretion 
\citep[e.g.][]{keres05,brooks09} are counted as ``smooth accretion'', 
which is dominated by cold flows at high redshift, and at all 
times that the virial mass is less than 10$^{11.5}$M$_\odot$. Yet, 
significant amounts of this gas which is classified as cold flow 
accretion can be involved in significant, and even major, gas-rich 
mergers subsequent to being accreted. The importance of gas-rich mergers 
is thus severely underestimated if these studies are superficially 
interpreted. Of course this goes both ways: the 49\% of the thick disc 
that we have classified as ``in situ gas mergers'' were formed form gas 
that first accreted to the disc region  smoothly.
 
Thin disc stars are dominated by smoothly accreted gas. We note that a not 
insignificant amount of gas that feeds the thin disc does come from 
gas-rich mergers. Much of this is recycled to the disc via the hot halo, 
after being ejected from the star forming regions of the galaxy during 
starbursts. This large scale galactic fountain process allows the 
recycled gas to gain angular momentum (B12) and aids in the suppression 
of the ubiquitous G-dwarf problem (Pilkington et al. 2012 in prep).

\subsection{Thick Disc to Thin Disc Transition}
There is no \it significant \rm ``event'' occurring $\sim$6$-$7~Gyr ago 
that we can associate with the separation of the thin and thick discs. 
 However,  the galaxy's star formation rate was 
relatively high between $\sim$7$-$10~Gyr ago, and decreased somewhat 
near the time of the transition to thin disc star formation. 
In Figure~\ref{sfr_rad} we plot the star formation history of all stars 
of the simulated galaxy as the black dashed line, plotting the fraction of 
stars born per Gyr in bins of 0.5~Gyr. We also show the star formation 
history of stars in the disc region ($7<R_{xy}<8$\,kpc and $|z|<0.5$\,kpc,  black line), and the star formation 
history of disc stars according to different formation radii : 
3$<$R$_{\rm Form}$$<$5\,kpc (red), 5$<$R$_{\rm Form}$$<$7\,kpc (light blue), and 
7$<$R$_{\rm Form}$$<$9\,kpc (blue). R$_{\rm Form}$,  calculated using physical units, is the distance from the centre of the galaxy at which the star forms.
This demonstrates the inside-out nature of 
the disc growth, which will have consequences for the abundances that we 
discuss below. 
The red  line of Figure~\ref{sfr_rad} is also interesting in this respect. The 
stars in the solar annulus at $z$=0 that formed in the region 
3$<$R$_{\rm Form}$$<$5\,kpc (the region where thick disc stars predominantly 
formed) also has a star formation rate that declined around the time of 
transition to thin disc star formation. Stars formed in regions further 
out have more a gradual build-up of star formation, meaning a longer 
enrichment timescale.

 \begin{table}
\begin{tabular}{ccccc}
\hline
&&& Metal Poor& Metal Rich\\  
&Total& $\alpha$-Old& $\alpha$-Young &$\alpha$-Young \\   
\hline
h$_l$ &2.73&2.31&4.07&2.74\\ 
\hline
h$_z$&0.7&0.9&0.7 &0.5\\ 
\hline
\end{tabular}
\label{struct}
\caption{The scale-length (h$_l$) and scale-height (h$_z$) of stars as selected by abundances, 
defining $\alpha$-Old as those above the dashed line in Fig.~3, and  $\alpha$-Young as those below the line. $\alpha$-Young are then split using a simple metallicity cut at [Fe/H]=$-0.68$. Metal Poor $\alpha$-Young have [Fe/H]$<-0.68$ and Metal Rich as $\alpha$-Young have [Fe/H]$>-0.68$.
Stars are from the region $4<{\rm R}_{xy}<12$\,kpc and $0<|{\rm z}|<4$\,kpc, while scale heights use stars in the annulus $7<{\rm R}_{xy}<8$\,kpc and are measured out to 4\,kpc.}
\end{table}

\section{Structure of Abundance selected Populations}
Here we mimic \cite{bovy12b}, and analyse the scale-length (h$_l$) and scale-height (h$_z$) of stars as selected by abundances. We use our thick and thin disc separation as shown in  our Fig.~3 to delineate the $\alpha$-old population as those above the dashed line, and  $\alpha$-young as those below  (using the nomenculture used in  \citealt{bovy12b}). $\alpha$-young are then split using a simple metallicity cut\footnote{results are not sensitive to this exact value} at [Fe/H]=$-0.68$. Metal poor $\alpha$-young  (MP$\alpha$-Y) have [Fe/H]$<-0.68$ and metal rich  $\alpha$-young (MR$\alpha$-Y) have [Fe/H]$>-0.68$. In this section we have extended the region of stars beyond the solar annulus in order to measure these properties, but the metallicity cuts remain valid separators of the populations.  
Stars taken  are from the region $4<{\rm R}_{xy}<12$\,kpc and $0<|{\rm z}|<4$\,kpc, while scale heights use stars in the annulus $7<{\rm R}_{xy}<8$\,kpc and are measured out to 4\,kpc. The scale-lengths and scale-heights of the three populations are listed in Table~2.

\begin{figure}
\hspace{-.7cm}  \includegraphics[width=.37\textheight]{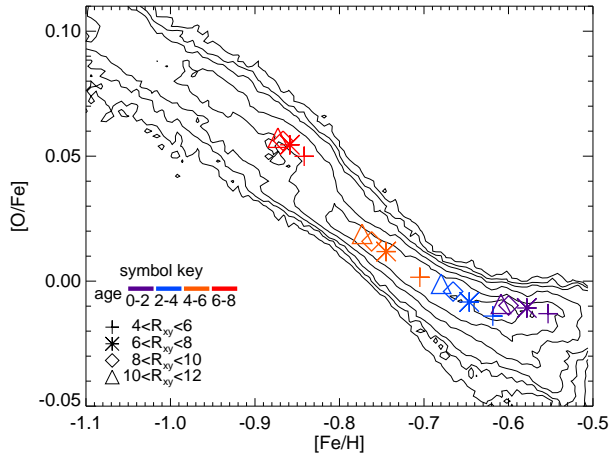}
\caption{The evolution through abundance space of stars in different radial regions. The symbols represent median values of  different regions in the disk, 
with ${\rm R}_{xy}$ (kpc) 4-6\,kpc ('+' sign), 6-8\,kpc (star), 8-10\,kpc (diamond), 10-12\,kpc (triangle). At any given time, the inner region is more chemically evolved than the  
outer region. This explains the longer scale length of the less evolved metal poor $\alpha$-young population  (MP$\alpha$-Y) compared with the metal rich  $\alpha$-young population (MR$\alpha$-Y).}
\label{alphaevolve}
\end{figure}

As already noted, our $\alpha$-old (thick disc) population has a relatively short scale-length (2.31\,kpc) and large scale-height (0.9\,kpc) compared to the total population which has values of 2.73\,kpc and 0.7\,kpc respectively. The features of the $\alpha$-young are a little less simple, and interestingly match very well the findings in \cite{bovy12b}. Firstly looking looking at scale-heights within the solar annulus,  the MP$\alpha$-Y has a significantly larger scale-height than the MR$\alpha$-Y population. This simply reflects their average age, with metal rich stars being generally younger. Yet
 the MP$\alpha$-Y population has a significantly larger scale-length than the MR$\alpha$-Y population.
 This is the result  of the difference in chemical abundance timescales between the inner and outer disks, relating to inside out disc growth. 
The inner regions are the most chemically evolved, meaning that  there remains greater relative numbers of  un-evolved MP$\alpha$-Y stars out further in the disk which results in their population having a longer scale length. Figure~\ref{alphaevolve} highlights  this. The symbols represent median values of  different regions in the disk, 
with ${\rm R}_{xy}$  4-6\,kpc (+ sign), 6-8\,kpc (star), 8-10\,kpc (diamond), 10-12\,kpc (triangle). At any given time, the inner region is more chemically evolved than regions further out. 

If migration is efficient, stars migrating to the outer disk will pollute the gas there with SN Ia, i.e. reduce the O/Fe ratio in the outer regions (see Figure 14 in \citealt{loebman11}). Thus, the degree of  migration  may be refleceted in the observed  difference in scale-length between   MP$\alpha$-Y  and  MR$\alpha$-Y populations.

\section{Migration}\label{migration}
\begin{figure}
\hspace{-.4cm}  \includegraphics[height=.28\textheight]{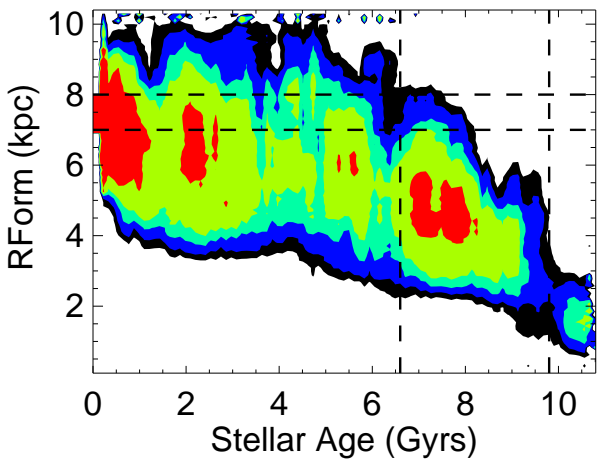}
\caption{The evolution of the distribution of formation radii for stars 
that end up in the solar annulus: $7<R_{xy}<8$, $|z|<0.5$.}
\label{migrate}
\end{figure}

Migration proves to be another key process affecting the distribution of galactic components.  Figure~\ref{migrate}  shows where stars formed that end up in the stellar annulus.  The stellar annulus is demarked by the horizontal dashed lines.  The dotted vertical lines show the divisions between age defined thin disc, thick disc and halo populations.  The mean formation radius for thick disk stars is closer to the centre of the galaxy than the thin disc stars.  Clearly, thick 
disc stars are born closer to the centre of the galaxy than the later 
forming thin disc stars. We do not further examine the mode of migration here
- i.e., we do not distinguish the contributions of churning, blurring, 
scattering from the bar, or co-rotation scattering in this study. Briefly, we can compare  our plot and Figures~3\,\&\,5 of \cite{loebman11}: the degree of migration is larger than in their Figure~5, the case where little co-rotation scattering occurs due to weak spiral structure, but is lower than in Figure~3, where significant co-rotation scattering occurs.

\begin{figure}
\hspace{-.4cm}  
\includegraphics[height=.25\textheight]{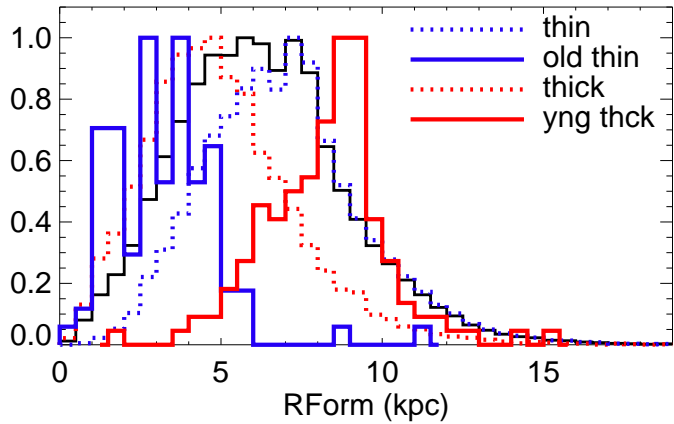}
\caption{The thin black line shows the distribution of radii of star 
formation of solar annulus stars, with the distribution of thin (blue 
dotted) and thick (red dotted) overplotted. The blue and red lines show 
the extremely old (age$>$7.8Gyr) thin and young (age $<$5.2~Gyr) thick 
disc stars.}
\label{migrate_extremes}
\end{figure}

Figure~\ref{migrate_extremes} shows a component by component analysis of migration.  For this analysis, we revert to selecting the components based on metallicity.  The solid black line shows the total distribution of formation radius.  The red line shows the thick disk and the blue line shows the thin disk distributions.  

We separate outlier stars out of the thick and thin disk populations, to try to understand why some young stars share abundances with the thick disc population, while some old stars share abundances with the thin disc.  The young (age $<$ 5.2 Gyr) thick disk population is shown as the dotted red line, and the old (age $>$ 7.8 Gyr) thin disc population is shown as the dotted blue line.  Clearly, old thin 
disc stars were born closer to the centre of the galaxy than the general 
disc population, and at even smaller radii than the thick disc 
population. This is explained by the fact that the inner region is more 
chemically evolved because the disc grows inside-out 
\citep[see also][]{pilkingtonfew12}, so the [$\alpha$/Fe] ratio becomes lower at an 
earlier time in the innermost region than in regions further out. Thus, 
old stars born in this inner region have thin disc abundances. 
Contrastingly, young thick disc stars (based on abundances) migrated 
from outer regions, which were relatively un-evolved chemically. This is 
similar to findings in the models of \citep{sb09b}

We also examine the dispersion of stars in these populations, and find 
that, for the old thin disc $\sigma_{\rm W}$=31~km/s, comparable to 
other stars of similar age.  $\sigma_{\rm W}$ of young thick disc stars 
(migrated from outer regions) is $\sim$36~km/s, similar to the rest of 
the thick disc population. We compare with young stars that have 
migrated from the inner regions (R$_{\rm Form}$$<$5~kpc - i.e., the region 
from which thick disc stars predominantly migrated) and find 
$\sigma_{\rm W}$=25~km/s. This indicates that migration alone will not 
build a thick disc in our simulation, and highlights the importance that 
the original old disc is born relatively hot, as shown in 
Figure~\ref{sigma_radius}.

\section{Discussion}\label{discussion}
\subsection{Milky Way Comparisons}

We do not want to over-emphasise any comparison to the Milky Way that 
can be drawn from our study, and remind the reader again that the 
simulated galaxy is significantly lower mass, and will have a different 
merger history.  We do want to emphasise that this simulated galaxy has 
a low mass, low metallicity, stellar halo. The majority of previous 
simulations that have analysed the origins of stellar components have 
had overly massive spheroids \citep[e.g.][]{abadi03b,zolotov09,kobayashi11,tissera12} (from 
the accretion of an excess of stars) and suffered from an inability to 
match the empirical stellar mass-halo mass relation 
\citep[e.g.][]{piontek11}.  The mass and metallicity of the stellar halo 
provides strict constraints on galaxy formation simulations 
(\citealt{brook04b}; \citealt{okamoto05}). We showed that we were able separate thin and thick 
disc stars based on abundances (and age), in a similar manner to 
\cite{navarro11} for the Milky Way. The qualitative similarities between 
the simulation and the Milky Way of having two tracks for the two 
populations through the [O/Fe], [Fe/H] plane are indeed interesting, and 
we attribute the change from thick disc to thin disc star formation to a 
drop in the star formation rate, associated with the end of the rapid 
gas accretion phase \citep{keres05} and the end of the gas rich merger 
epoch.  The {\it degree} of separation of 
these tracks may be an instructive diagnostic: the two 
tracks in the simulation are closer together (separated by 
$\sim$0.01-0.05~dex in [O/Fe]) than the Milky Way tracks (separated by 
$\sim$0.2~dex in [O/Fe] \citealt{bensby05,reddy06,fuhrmann08}), which has been interpreted by chemical evolution models as indicating that the Milky Way 
experienced a {\it hiatus} in star formation \citep[e.g.][]{chiappini01,fenner03}.   \cite{sb09b} attain  bimodal distributions in [O/Fe] as a  consequence of their  assumptions about star-formation rates and metal enrichment,  without a star formation hiatus (their star formation is assumed to exponentially decay). Yet we remind the reader that the degree of offset in the observed relations in  Milky Way may, at least  partially, be due to the   kinematic selection which is biased  toward thick-disc stars.

\subsection{Components}
We have made a component analysis of a simulated disc galaxy. We categorise stars 
as belonging to the thin disc, thick disc, and halo by simple cuts in 
abundances of [O/Fe] and [Fe/H] in a manner similar to that done for 
observed stars in \cite{navarro11}. In qualitative sense, the properties 
of the components are reasonable analogues of the Milky Way components, 
in terms of velocity dispersion, rotation velocities, and metallicy.  We 
trace the origin of gas from which stars in the different components 
form:\\

\noindent{\bf Halo:} In the solar neighbourhood, most stars form {\it in 
situ} but are knocked into the halo during the merger epoch.  This is 
quantitatively higher than the fraction of {\it in situ} halo stars as 
suggested in \citep{zolotov09}, who predicted a significant fraction of 
halo stars are formed \it in situ\rm, but found that accretion remains 
the dominant source of halo stars. We attribute this difference to the 
low stellar mass fractions of the accreted protogalaxies/satellites in 
our simulations, which better match empirical relations between stellar 
and halo mass (\citealt[]{brook12b}). This also results in the low mass, low-metallicity halo 
of the current simulation. By contrast, using physical recipes similar 
to those used in \cite{zolotov09} has been shown to result in 
significantly too many stars compared to the empirical relation between 
stellar and total halo masses \citep{sawala11} in low mass galaxies, 
exaggerating the number of accreted stars and resulting in spheroids 
that are more massive and more metal-rich than observed in the Milky 
Way.  In our simulated galaxy there is still a significant fraction 
($\sim$17-31\%) of local halo stars that are directly accreted from 
smaller proto-galaxies/satellites.\\

\noindent{\bf Thick Disc:} Thick disc stars all form  {\it in situ}. 
Again this is in contrast to numerous studies of simulated Milky Way 
mass galaxies, which claim significant numbers of accreted thick disc 
stars \citep[e.g.][]{abadi03b,kobayashi11,tissera12}. We note that those previous 
studies have stellar spheroids that are significantly more massive than 
observed in the Milky Way. Significant numbers of accreted thick disc 
stars have not been shown to form without corresponding massive 
spheroids also being accreted \citep[e.g.][]{abadi06,tissera12}. In our 
simulated galaxy, around 77\% of thick disc stars form from gas that is 
directly accreted to the central galaxy, while 23\% is accreted as 
clumpy gas - i.e., gas which is part of an accreted 
proto-galaxy/satellite. Yet, most of the thick disc forming gas is 
accreted prior to the end of the `merger epoch', meaning that most 
($\sim$80\%) of thick disc gas is involved in a gas-rich merger at high 
redshift. Thick disc stars form as the gas settles into a disc at the 
end of the merger epoch. The thick disc has a relatively short radial 
scalelength and is kinematically `hot' as it forms. \\

\noindent{\bf Thin Disc:} The thin disc stars form primarily from gas 
that is smoothly accreted to the central galaxy subsequent to the end of 
the merger epoch. A significant number of thin disc stars form from gas 
accreted during the merger epoch, either directly to the central galaxy 
or as part of an accreted proto-galaxy, which is cycled through the 
galaxy's hot corona in a large scale galactic fountain 
\cite[see][]{brook12a}.\\

\noindent{\bf Transitions:} In our simulation, the end of the merger 
epoch marks the transition from halo star formation to thick disc star 
formation.  No ``event'' marks the 
transition from thick to thin disc in our simulation. We do note that 
the thick disc formation time is longer in our simulation than most 
estimates of the Milky Way thick disc ($\sim$3.5\,Gyr compared to 1-2 
Gyrs \citealt[e.g.][]{fuhrmann08,smiljanic09}, although 
\citealt{bensby07} claims a timescale of 3\,Gyr). This may be due to the 
lower mass of our simulated galaxy, which would result in a longer time 
for the thick disc to settle down. This would be consistent with 
observations which indicate that thick discs are more prominent in lower 
mass disc galaxies \citep{yoachim06}. Alternatively, it may of course be 
a problem with our model.\\

\noindent{\bf Inside-out Chemical  Evolution:} 
The thick disc (with an $\alpha$-old population, i.e. stars above the dashed line in Fig. 3) has a scale length of 2.31\,kpc at $z=0$. The inside-out growth of the thin disc ($\alpha$-young stars, below the dashed line in Fig. 3) results in  differences in chemical abundance timescales between the inner and outer disks. The inner regions are the most chemically evolved, meaning that  there remains greater relative numbers of  un-evolved metal poor $\alpha$-young stars out further in the disk, which results in that population having a longer scale length (4.07\,kpc) compared to the metal-rich  $\alpha$-young stars (2.74\,kpc). \\

\noindent{\bf Migration:} The stars classified as thick disc in our 
model preferentially were born closer to the centre of the galaxy, and 
migrated to the solar neighbourhood. Our model is fully cosmological, 
yet it is chosen to have a quiescent merger period at late ($z<1.5$) 
times. At these times, it thus resembles the models of 
\cite{loebman11}, and it is interesting that their collapse model also 
initially forms a relatively short, kinematically hot, disc at high 
redshift before the thin disc formation begins. The dissipative nature 
of the gas rich merger epoch in our case and the collapse of a rotating 
sphere in their case (and in the study of \citealt{samland04}) has 
resulted in information of the early differences between the models being lost, at least in 
kinematics alone. \cite{loebman11} also found that simple cuts in abundance space could result in
populations with thick and thin disc properties.
Detailed chemical ``fingerprints'' that will become 
available in surveys such as GALAH may combine with GAIA kinematic data 
to unravel star formation cites \citep{freeman02,desilva06}.  We also 
note that the initial conditions imposed in the dynamical model of 
\cite{sb09b} consist of a relatively hot and short disc. Our 
simulations suggest that migration is part of the story of thick disc 
formation, but not necessarily to the exclusion of other theories. The 
recent paper of \citep{liu12} finds evidence for both migration and gas 
rich mergers in the relationships between eccentricities and abundances.

\section*{Acknowledgments}
BKG and CBB acknowledge the support of the UK's Science \& Technology 
Facilities Council (ST/F002432/1 \& ST/H00260X/1). BKG and KP 
acknowledge the generous visitor support provided by Saint Mary's 
University and Monash University. TRQ was supported by NSF Grant AST- 
0908499. We thank the DEISA consortium, co-funded through EU FP6 project 
RI-031513 and the FP7 project RI-222919, for support within the DEISA 
Extreme Computing Initiative, the UK's National Cosmology Supercomputer 
(COSMOS), and the University of Central Lancashire's High Performance 
Computing Facility. C.B. Brook and A.V. Macci\`o acknowledge funding by Sonderforschungsbereich SFB 881 ``The Milky Way System" (subproject A1) of the German Research Foundation (DFG).

\bibliographystyle{mn2e}
\bibliography{brook}

\begin{thebibliography}{78}
\expandafter\ifx\csname natexlab\endcsname\relax\def\natexlab#1{#1}\fi

\bibitem[{{Abadi} {et~al}\mbox{.}(2006){Abadi}, {Navarro}, \&
  {Steinmetz}}]{abadi06}
{Abadi} M.~G., {Navarro} J.~F., {Steinmetz} M., 2006, MNRAS, 365, 747

\bibitem[{{Abadi} {et~al}\mbox{.}(2003){Abadi}, {Navarro}, {Steinmetz}, \&
  {Eke}}]{abadi03b}
{Abadi} M.~G., {Navarro} J.~F., {Steinmetz} M., {Eke} V.~R., 2003, ApJ, 597, 21

\bibitem[{{Agertz} {et~al}\mbox{.}(2009){Agertz}, {Teyssier}, \&
  {Moore}}]{agertz09}
{Agertz} O., {Teyssier} R., {Moore} B., 2009, MNRAS, 397, L64

\bibitem[{{Assmann} {et~al}\mbox{.}(2011){Assmann}, {Fellhauer}, {Kroupa},
  {Bruens}, \& {Smith}}]{assmann11}
{Assmann} P., {Fellhauer} M., {Kroupa} P., {Bruens} R.~C., {Smith} R., 2011,
  ArXiv e-prints

\bibitem[{{Bekki} \& {Tsujimoto}(2011)}]{bekki11}
{Bekki} K., {Tsujimoto} T., 2011, ArXiv e-prints

\bibitem[{{Bensby} \& {Feltzing}(2011)}]{bensbyfeltzing11}
{Bensby} T., {Feltzing} S., 2011, ArXiv e-prints

\bibitem[{{Bensby} {et~al}\mbox{.}(2003){Bensby}, {Feltzing}, \&
  {Lundstr{\"o}m}}]{bensby03}
{Bensby} T., {Feltzing} S., {Lundstr{\"o}m} I., 2003, AAP, 410, 527

\bibitem[{{Bensby} {et~al}\mbox{.}(2005){Bensby}, {Feltzing}, {Lundstr{\"o}m},
  \& {Ilyin}}]{bensby05}
{Bensby} T., {Feltzing} S., {Lundstr{\"o}m} I., {Ilyin} I., 2005, AAP, 433, 185

\bibitem[{{Bensby} {et~al}\mbox{.}(2007){Bensby}, {Zenn}, {Oey}, \&
  {Feltzing}}]{bensby07}
{Bensby} T., {Zenn} A.~R., {Oey} M.~S., {Feltzing} S., 2007, ApJL, 663, L13

\bibitem[{{Bournaud} {et~al}\mbox{.}(2007){Bournaud}, {Elmegreen}, \&
  {Elmegreen}}]{bournaud07}
{Bournaud} F., {Elmegreen} B.~G., {Elmegreen} D.~M., 2007, ApJ, 670, 237

\bibitem[{{Bovy} {et~al}\mbox{.}(2012){Bovy}, {Rix}, \& {Hogg}}]{bovy12a}
{Bovy} J., {Rix} H.-W., {Hogg} D.~W., 2012, ApJ, 751, 131

\bibitem[{{Bovy} {et~al}\mbox{.}(2011){Bovy}, {Rix}, {Liu}, {Hogg}, {Beers}, \&
  {Lee}}]{bovy12b}
{Bovy} J., {Rix} H.-W., {Liu} C., {Hogg} D.~W., {Beers} T.~C., {Lee} Y.~S.,
  2011, ArXiv e-prints

\bibitem[{{Brook} {et~al}\mbox{.}(2004{\natexlab{a}}){Brook}, {Kawata},
  {Gibson}, \& {Flynn}}]{brook04b}
{Brook} C.~B., {Kawata} D., {Gibson} B.~K., {Flynn} C., 2004{\natexlab{a}},
  MNRAS, 349, 52

\bibitem[{{Brook} {et~al}\mbox{.}(2004{\natexlab{b}}){Brook}, {Kawata},
  {Gibson}, \& {Freeman}}]{brook04a}
{Brook} C.~B., {Kawata} D., {Gibson} B.~K., {Freeman} K.~C.,
  2004{\natexlab{b}}, ApJ, 612, 894

\bibitem[{{Brook} {et~al}\mbox{.}(2012{\natexlab{a}}){Brook}, {Stinson},
  {Gibson}, {Ro{\v s}kar}, {Wadsley}, \& {Quinn}}]{brook12a}
{Brook} C.~B., {Stinson} G., {Gibson} B.~K., {Ro{\v s}kar} R., {Wadsley} J.,
  {Quinn} T., 2012{\natexlab{a}}, MNRAS, 419, 771

\bibitem[{{Brook} {et~al}\mbox{.}(2012{\natexlab{b}}){Brook}, {Stinson},
  {Gibson}, {Wadsley}, \& {Quinn}}]{brook12b}
{Brook} C.~B., {Stinson} G., {Gibson} B.~K., {Wadsley} J., {Quinn} T.,
  2012{\natexlab{b}}, ArXiv e-prints

\bibitem[{{Brooks} {et~al}\mbox{.}(2009){Brooks}, {Governato}, {Quinn},
  {Brook}, \& {Wadsley}}]{brooks09}
{Brooks} A.~M., {Governato} F., {Quinn} T., {Brook} C.~B., {Wadsley} J., 2009,
  ApJ, 694, 396

\bibitem[{{Ceverino} {et~al}\mbox{.}(2010){Ceverino}, {Dekel}, \&
  {Bournaud}}]{ceverino10}
{Ceverino} D., {Dekel} A., {Bournaud} F., 2010, MNRAS, 404, 2151

\bibitem[{{Chiappini} {et~al}\mbox{.}(2001){Chiappini}, {Matteucci}, \&
  {Romano}}]{chiappini01}
{Chiappini} C., {Matteucci} F., {Romano} D., 2001, ApJ, 554, 1044

\bibitem[{{Chiba} \& {Beers}(2000)}]{chibabeers00}
{Chiba} M., {Beers} T.~C., 2000, AJ, 119, 2843

\bibitem[{{Dalcanton} \& {Bernstein}(2002)}]{dalcanton02}
{Dalcanton} J.~J., {Bernstein} R.~A., 2002, AJ, 124, 1328

\bibitem[{{De Silva} {et~al}\mbox{.}(2006){De Silva}, {Sneden}, {Paulson},
  {Asplund}, {Bland-Hawthorn}, {Bessell}, \& {Freeman}}]{desilva06}
{De Silva} G.~M., {Sneden} C., {Paulson} D.~B., {Asplund} M., {Bland-Hawthorn}
  J., {Bessell} M.~S., {Freeman} K.~C., 2006, AJ, 131, 455

\bibitem[{{Dom{\'e}nech-Moral} {et~al}\mbox{.}(2012){Dom{\'e}nech-Moral},
  {Mart{\'{\i}}nez-Serrano}, {Dom{\'{\i}}nguez-Tenreiro}, \&
  {Serna}}]{domenech12}
{Dom{\'e}nech-Moral} M., {Mart{\'{\i}}nez-Serrano} F.~J.,
  {Dom{\'{\i}}nguez-Tenreiro} R., {Serna} A., 2012, ArXiv e-prints

\bibitem[{{Fenner} \& {Gibson}(2003)}]{fenner03}
{Fenner} Y., {Gibson} B.~K., 2003, PASA, 20, 189

\bibitem[{{Freeman} \& {Bland-Hawthorn}(2002)}]{freeman02}
{Freeman} K., {Bland-Hawthorn} J., 2002, ARAA, 40, 487

\bibitem[{{Freyer} {et~al}\mbox{.}(2006){Freyer}, {Hensler}, \&
  {Yorke}}]{freyer06}
{Freyer} T., {Hensler} G., {Yorke} H.~W., 2006, ApJ, 638, 262

\bibitem[{{Fuhrmann}(1998)}]{fuhrmann98}
{Fuhrmann} K., 1998, AAP, 338, 161

\bibitem[{{Fuhrmann}(2008)}]{fuhrmann08}
{Fuhrmann} K., 2008, MNRAS, 384, 173

\bibitem[{{Gilmore} \& {Reid}(1983)}]{gilmore83}
{Gilmore} G., {Reid} N., 1983, MNRAS, 202, 1025

\bibitem[{{Governato} {et~al}\mbox{.}(2007){Governato}, {Willman}, {Mayer},
  {Brooks}, {Stinson}, {Valenzuela}, {Wadsley}, \& {Quinn}}]{g07}
{Governato} F., {Willman} B., {Mayer} L., {Brooks} A., {Stinson} G.,
  {Valenzuela} O., {Wadsley} J., {Quinn} T., 2007, MNRAS, 374, 1479

\bibitem[{{Guedes} {et~al}\mbox{.}(2011){Guedes}, {Callegari}, {Madau}, \&
  {Mayer}}]{guedes11}
{Guedes} J., {Callegari} S., {Madau} P., {Mayer} L., 2011, ApJ, 742, 76

\bibitem[{{Hammer} {et~al}\mbox{.}(2007){Hammer}, {Puech}, {Chemin}, {Flores},
  \& {Lehnert}}]{hammer07}
{Hammer} F., {Puech} M., {Chemin} L., {Flores} H., {Lehnert} M.~D., 2007, ApJ,
  662, 322

\bibitem[{{Hayashi} \& {Chiba}(2006)}]{hayashichiba06}
{Hayashi} H., {Chiba} M., 2006, PASJ, 58, 835

\bibitem[{{House} {et~al}\mbox{.}(2011){House}, {Brook}, {Gibson},
  {Sanchez-Blazquez}, {Courty}, {Few}, {Governato}, {Kawata}, {Roskar},
  {Steinmetz}, {Stinson}, \& {Teyssier}}]{house11}
{House} E. {et~al.}, 2011, ArXiv e-prints

\bibitem[{{Ivezi{\'c}} {et~al}\mbox{.}(2008){Ivezi{\'c}}, {Sesar}, {Juri{\'c}},
  {Newberg}, {Beers}, {Allende Prieto}, {Long}, {Nitta}, {Snedden}, {Lee},
  {Harris}, {Brinkmann}, {Schneider}, \& {York}}]{ivezic08}
{Ivezi{\'c}} {\v Z}. {et~al.}, 2008, ApJ, 684, 287

\bibitem[{{Juri{\'c}} {et~al}\mbox{.}(2008){Juri{\'c}}, {Ivezi{\'c}}, {Brooks},
  {Lupton}, {Schlegel}, {Finkbeiner}, {Padmanabhan}, {Bond}, {Sesar},
  {Rockosi}, {Knapp}, {Gunn}, {Sumi}, {Schneider}, {Barentine}, {Brewington},
  \& {Brinkmann}}]{juric08}
{Juri{\'c}} M. {et~al.}, 2008, ApJ, 673, 864

\bibitem[{{Katz}(1992)}]{katz92}
{Katz} N., 1992, ApJ, 391, 502

\bibitem[{{Kazantzidis} {et~al}\mbox{.}(2008){Kazantzidis}, {Bullock},
  {Zentner}, {Kravtsov}, \& {Moustakas}}]{stelios08}
{Kazantzidis} S., {Bullock} J.~S., {Zentner} A.~R., {Kravtsov} A.~V.,
  {Moustakas} L.~A., 2008, ApJ, 688, 254

\bibitem[{{Kere{\v s}} {et~al}\mbox{.}(2005){Kere{\v s}}, {Katz}, {Weinberg},
  \& {Dav{\'e}}}]{keres05}
{Kere{\v s}} D., {Katz} N., {Weinberg} D.~H., {Dav{\'e}} R., 2005, MNRAS, 363,
  2

\bibitem[{{Kobayashi} \& {Nakasato}(2011)}]{kobayashi11}
{Kobayashi} C., {Nakasato} N., 2011, ApJ, 729, 16

\bibitem[{{Kroupa}(2002)}]{kroupa02a}
{Kroupa} P., 2002, MNRAS, 330, 707

\bibitem[{{Liu} \& {van de Ven}(2012)}]{liu12}
{Liu} C., {van de Ven} G., 2012, ArXiv e-prints

\bibitem[{{Loebman} {et~al}\mbox{.}(2011){Loebman}, {Ro{\v s}kar},
  {Debattista}, {Ivezi{\'c}}, {Quinn}, \& {Wadsley}}]{loebman11}
{Loebman} S.~R., {Ro{\v s}kar} R., {Debattista} V.~P., {Ivezi{\'c}} {\v Z}.,
  {Quinn} T.~R., {Wadsley} J., 2011, ApJ, 737, 8

\bibitem[{{Majewski}(1993)}]{majewski93}
{Majewski} S.~R., 1993, ARA\&A, 31, 575

\bibitem[{{Martig} {et~al}\mbox{.}(2012){Martig}, {Bournaud}, {Croton},
  {Dekel}, \& {Teyssier}}]{martig12}
{Martig} M., {Bournaud} F., {Croton} D.~J., {Dekel} A., {Teyssier} R., 2012,
  ArXiv e-prints

\bibitem[{{Navarro} {et~al}\mbox{.}(2011){Navarro}, {Abadi}, {Venn}, {Freeman},
  \& {Anguiano}}]{navarro11}
{Navarro} J.~F., {Abadi} M.~G., {Venn} K.~A., {Freeman} K.~C., {Anguiano} B.,
  2011, MNRAS, 412, 1203

\bibitem[{{Noguchi}(1999)}]{noguchi99}
{Noguchi} M., 1999, ApJ, 514, 77

\bibitem[{{Nomoto} {et~al}\mbox{.}(1997){Nomoto}, {Iwamoto}, {Nakasato},
  {Thielemann}, {Brachwitz}, {Tsujimoto}, {Kubo}, \& {Kishimoto}}]{nomoto97}
{Nomoto} K., {Iwamoto} K., {Nakasato} N., {Thielemann} F.-K., {Brachwitz} F.,
  {Tsujimoto} T., {Kubo} Y., {Kishimoto} N., 1997, Nuclear Physics A, 621, 467

\bibitem[{{Nordstr{\"o}m} {et~al}\mbox{.}(2004){Nordstr{\"o}m}, {Mayor},
  {Andersen}, {Holmberg}, {Pont}, {J{\o}rgensen}, {Olsen}, {Udry}, \&
  {Mowlavi}}]{nordstrom04}
{Nordstr{\"o}m} B. {et~al.}, 2004, AAP, 418, 989

\bibitem[{{Okamoto} {et~al}\mbox{.}(2005){Okamoto}, {Eke}, {Frenk}, \&
  {Jenkins}}]{okamoto05}
{Okamoto} T., {Eke} V.~R., {Frenk} C.~S., {Jenkins} A., 2005, MNRAS, 363, 1299

\bibitem[{{Pilkington} {et~al}\mbox{.}(2012){Pilkington}, {Few}, {Gibson},
  {Calura}, {Michel-Dansac}, {Thacker}, {Molla}, {Matteucci}, {Rahimi},
  {Kawata}, {Kobayashi}, {Brook}, {Stinson}, {Couchman}, {Bailin}, \&
  {Wadsley}}]{pilkingtonfew12}
{Pilkington} K. {et~al.}, 2012, ArXiv e-prints

\bibitem[{{Piontek} \& {Steinmetz}(2011)}]{piontek11}
{Piontek} F., {Steinmetz} M., 2011, MNRAS, 410, 2625

\bibitem[{{Qu} {et~al}\mbox{.}(2011){Qu}, {Di Matteo}, {Lehnert}, \& {van
  Driel}}]{qu11}
{Qu} Y., {Di Matteo} P., {Lehnert} M.~D., {van Driel} W., 2011, A\&A, 530, A10

\bibitem[{{Quillen} \& {Garnett}(2001)}]{quillen01}
{Quillen} A.~C., {Garnett} D.~R., 2001, in Astronomical Society of the Pacific
  Conference Series, Vol. 230, Galaxy Disks and Disk Galaxies, {J.~G.~Funes \&
  E.~M.~Corsini}, ed., pp. 87--88

\bibitem[{{Quinn} {et~al}\mbox{.}(1993){Quinn}, {Hernquist}, \&
  {Fullagar}}]{quinn93}
{Quinn} P.~J., {Hernquist} L., {Fullagar} D.~P., 1993, ApJ, 403, 74

\bibitem[{{Reddy} {et~al}\mbox{.}(2006){Reddy}, {Lambert}, \& {Allende
  Prieto}}]{reddy06}
{Reddy} B.~E., {Lambert} D.~L., {Allende Prieto} C., 2006, MNRAS, 367, 1329

\bibitem[{{Robertson} \& {Kravtsov}(2008)}]{robertson08}
{Robertson} B.~E., {Kravtsov} A.~V., 2008, ApJ, 680, 1083

\bibitem[{{Ruchti} {et~al}\mbox{.}(2010){Ruchti}, {Fulbright}, {Wyse},
  {Navarro}, {Parker}, {Reid}, {Seabroke}, {Siebert}, {Siviero}, {Steinmetz},
  {Watson}, {Williams}, \& {Zwitter}}]{ruchti10}
{Ruchti} G.~R. {et~al.}, 2010, ApJL, 721, L92

\bibitem[{{Sales} {et~al}\mbox{.}(2011){Sales}, {Navarro}, {Theuns}, {Schaye},
  {White}, {Frenk}, {Crain}, \& {Dalla Vecchia}}]{sales12}
{Sales} L.~V., {Navarro} J.~F., {Theuns} T., {Schaye} J., {White} S.~D.~M.,
  {Frenk} C.~S., {Crain} R.~A., {Dalla Vecchia} C., 2011, ArXiv e-prints

\bibitem[{{Samland} \& {Gerhard}(2003)}]{samland04}
{Samland} M., {Gerhard} O.~E., 2003, AAP, 399, 961

\bibitem[{{Sawala} {et~al}\mbox{.}(2011){Sawala}, {Guo}, {Scannapieco},
  {Jenkins}, \& {White}}]{sawala11}
{Sawala} T., {Guo} Q., {Scannapieco} C., {Jenkins} A., {White} S., 2011, MNRAS,
  413, 659

\bibitem[{{Scannapieco} {et~al}\mbox{.}(2010){Scannapieco}, {Gadotti},
  {Jonsson}, \& {White}}]{scannapieco10}
{Scannapieco} C., {Gadotti} D.~A., {Jonsson} P., {White} S.~D.~M., 2010, MNRAS,
  407, L41

\bibitem[{{Scannapieco} {et~al}\mbox{.}(2011){Scannapieco}, {Wadepuhl},
  {Parry}, {Monaco}, {Murante}, {Okamoto}, {Quinn}, {Schaye}, {Stinson},
  {Theuns}, {Wadsley}, {White}, \& {Woods}}]{scannapieco12}
{Scannapieco} C. {et~al.}, 2011, ArXiv e-prints

\bibitem[{{Sch{\"o}nrich} \& {Binney}(2009)}]{sb09b}
{Sch{\"o}nrich} R., {Binney} J., 2009, MNRAS, 399, 1145

\bibitem[{{Shen} {et~al}\mbox{.}(2010){Shen}, {Wadsley}, \& {Stinson}}]{shen10}
{Shen} S., {Wadsley} J., {Stinson} G., 2010, MNRAS, 407, 1581

\bibitem[{{Smiljanic} {et~al}\mbox{.}(2009){Smiljanic}, {Pasquini},
  {Bonifacio}, {Galli}, {Gratton}, {Randich}, \& {Wolff}}]{smiljanic09}
{Smiljanic} R., {Pasquini} L., {Bonifacio} P., {Galli} D., {Gratton} R.~G.,
  {Randich} S., {Wolff} B., 2009, AAP, 499, 103

\bibitem[{{Soubiran} {et~al}\mbox{.}(2003){Soubiran}, {Bienaym{\'e}}, \&
  {Siebert}}]{soubiran03}
{Soubiran} C., {Bienaym{\'e}} O., {Siebert} A., 2003, AAP, 398, 141

\bibitem[{{Stinson} {et~al}\mbox{.}(2006){Stinson}, {Seth}, {Katz}, {Wadsley},
  {Governato}, \& {Quinn}}]{stinson06}
{Stinson} G., {Seth} A., {Katz} N., {Wadsley} J., {Governato} F., {Quinn} T.,
  2006, MNRAS, 373, 1074

\bibitem[{{Stinson} {et~al}\mbox{.}(2010){Stinson}, {Bailin}, {Couchman},
  {Wadsley}, {Shen}, {Nickerson}, {Brook}, \& {Quinn}}]{stinson10}
{Stinson} G.~S., {Bailin} J., {Couchman} H., {Wadsley} J., {Shen} S.,
  {Nickerson} S., {Brook} C., {Quinn} T., 2010, MNRAS, 408, 812

\bibitem[{{Tissera} {et~al}\mbox{.}(2011){Tissera}, {White}, \&
  {Scannapieco}}]{tissera12}
{Tissera} P.~B., {White} S.~D.~M., {Scannapieco} C., 2011, MNRAS, 2002

\bibitem[{{Torres}(2010)}]{torres10}
{Torres} G., 2010, AJ, 140, 1158

\bibitem[{{Villalobos} \& {Helmi}(2008)}]{villalobos08}
{Villalobos} {\'A}., {Helmi} A., 2008, MNRAS, 391, 1806

\bibitem[{{Wadsley} {et~al}\mbox{.}(2004){Wadsley}, {Stadel}, \&
  {Quinn}}]{wadsley04}
{Wadsley} J.~W., {Stadel} J., {Quinn} T., 2004, New Astronomy, 9, 137

\bibitem[{{Woosley} \& {Weaver}(1995)}]{woosley95}
{Woosley} S.~E., {Weaver} T.~A., 1995, ApJS, 101, 181

\bibitem[{{Wyse} {et~al}\mbox{.}(2006){Wyse}, {Gilmore}, {Norris}, {Wilkinson},
  {Kleyna}, {Koch}, {Evans}, \& {Grebel}}]{wyse06}
{Wyse} R.~F.~G., {Gilmore} G., {Norris} J.~E., {Wilkinson} M.~I., {Kleyna}
  J.~T., {Koch} A., {Evans} N.~W., {Grebel} E.~K., 2006, ApJL, 639, L13

\bibitem[{{Yoachim} \& {Dalcanton}(2005)}]{yoachim05}
{Yoachim} P., {Dalcanton} J.~J., 2005, ApJ, 624, 701

\bibitem[{{Yoachim} \& {Dalcanton}(2006)}]{yoachim06}
{Yoachim} P., {Dalcanton} J.~J., 2006, AJ, 131, 226

\bibitem[{{Zolotov} {et~al}\mbox{.}(2009){Zolotov}, {Friends}, {Friends},
  {Friends}, \& {Friends}}]{zolotov09}
{Zolotov} A., {Friends} S.~D.~M., {Friends} S.~D.~M., {Friends} S.~D.~M.,
  {Friends} S.~D.~M., 2009, MNRAS, 347, 556

\end{thebibliography}

\label{lastpage}

\end{document}